%% file: CKM_v5.tex
\documentclass[12pt]{article}
\usepackage{epsfig}
\usepackage[english]{babel}
\usepackage{graphicx}
\usepackage{hepparticles}
\usepackage{hepnicenames}
\usepackage{hepunits}
\usepackage[latin1]{inputenc} 
\usepackage[T1]{fontenc}
\usepackage{subfig}
\input econfmacros.tex 
\def\Title#1{\begin{center} {\Large {\bf #1} } \end{center}}

\begin{document}

\Title{Searches for LFV and LNV Decays at LHCb  }

\bigskip\bigskip


\begin{raggedright}  

{\it Marcin Chrz\k{a}szcz \index{Chrz\k{a}szcz, M.}, on behalf of the LHCb collaboration \\
H.Niedniczanski Institute of Nuclear Physics, Polish Academy of Science\\
ul. Radzikowskiego 152, 31-342 Cracow, Poland \\
{~}\\
Proceedings of CKM 2012, the 7th International Workshop on the CKM Unitarity Triangle, University of Cincinnati, USA, 28 September - 2 October 2012 
}

\bigskip\bigskip
\end{raggedright}

\begin{abstract}

The paper presents the latest progress on the searches for Lepton Number Violating (LNV) B Meson decays, the Lepton Flavour Violating (LFV) decay $\Ptau^- \to \Pmu^-\Pmu^-\Pmu^+$, and the Lepton and Baryon Number Violating (LNV and BNV) decays $\Ptau^- \to \mu^+ \mu^- \APproton$ and $\Ptau^- \to \mu^- \mu^- \Pproton$ at the LHCb. These searches have been performed at a hadron collider for the first time.   In the absence of signal we put upper limits, which are as follows: $\mathcal{B}(\tau^- \to \mu^- \mu^- \mu^+) < 6.3 \times 10^{-8}$, $\mathcal{B}(\Ptau^- \to \mu^- \mu^+ \APproton) < 3.4 \times 10^{-7}$, $\mathcal{B}(\Ptau^- \to \mu^- \mu^- \Pproton) < 4.4 \times 10^{-7}$ at $90\%$ CL.\\
Inclusion of charged conjugate processes are implied throughout this document.
\end{abstract}

\section{Introduction}

Lepton Flavour Violation has long been observed in the neutrino sector \cite{noscilation}; the phenomenon is known as neutrino oscillation. Charged LFV also arises in the Standard Model (SM) \cite{SM} from the neutrino mass terms via neutrino oscillation at loop level. This effect is suppressed by powers of $  {m^2_\nu}/{m_W^2}$, meaning the branching ratios are typically $< 10^{-54}$\cite{10_54} which is well below the observable level. However, many New Physics (NP) models (such as MSSM, R-party violating Supersymmetry and Littlest Higgs with T-parity \cite{jons}) predict LFV decays at much higher rates, significant enough to fall within the current experimental sensitivity in certain regions of the model parameter space. Whereas any direct observation of charged LFV will be a clear indication of NP, exclusion limits for LFV also serve as a powerful tool to exclude parameter spaces of NP models. \\

Many NP models also feature a wider class of processes: Lepton Number Violation (LNV) \cite{LNV} and Baryon Number Violation (BNV) which are strictly forbidden in the SM; most of them \cite{manymodels} predict $ \vert B - L \vert =0,2$ where $B$ is the baryon number and $L$ is the lepton number. Whereas BNV is highly speculated since it is one of the Sakharov \cite{saharow} conditions to explain the matter-antimatter asymmetry, LNV has not yet been observed despite over 70 years of extensive searches in neutrinoless double $\beta$ decay \cite{beta}.

\section{Majorana Neutrino and LNV in B Meson Sector}
Neutrinoless hadron LNV decays with like sign dilepton final state provide an important probe for the existence of Majorana neutrinos. In Majorana neutrino model, LNV B meson decays $\PB^- \to h^{+} l^- l^-$ are produced via two mechanisms at the lowest order, which involve either an on-shell or virtual Majorana neutrino (analogous to double $\beta$ decay). 
\begin{figure}[h] 
\begin{center} 
\mbox{
	\subfloat[{Virtual Majorana neutrino}]{\includegraphics[scale=0.85]{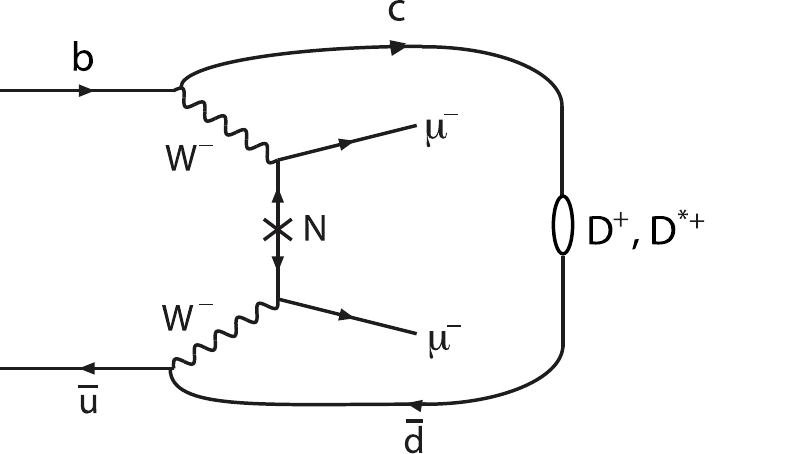}}
    \subfloat[{On shell Majorana neutrino}]{\includegraphics[scale=0.75]{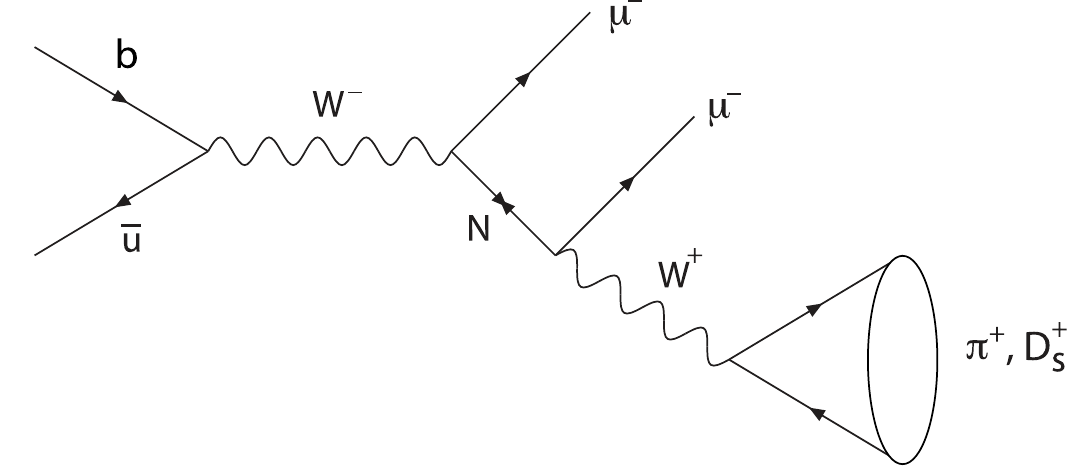}}
}
\caption{Example of the lowest order Feyman diagrams for $\PB$ meson LNV decays via Majorana neutrino.}
\end{center}
\end{figure}
On-shell production of $\PB^- \to \Ppi^+ \mu^- \mu^-$ and  $\PB^- \to  \PD^+_s \mu^- \mu^-$ enables us to probe Majorana neutrinos in mass range up to  $5140$ MeV/$c^2$. Beyond this mass the modes $\PB^- \to \PDplus \mu^- \mu^-$ and  $\PB^- \to  \PDplus^{\ast} \mu^- \mu^-$ are more restrictive.
\begin{figure}[h]
\begin{center} 
\mbox{
	{\includegraphics[scale=0.85]{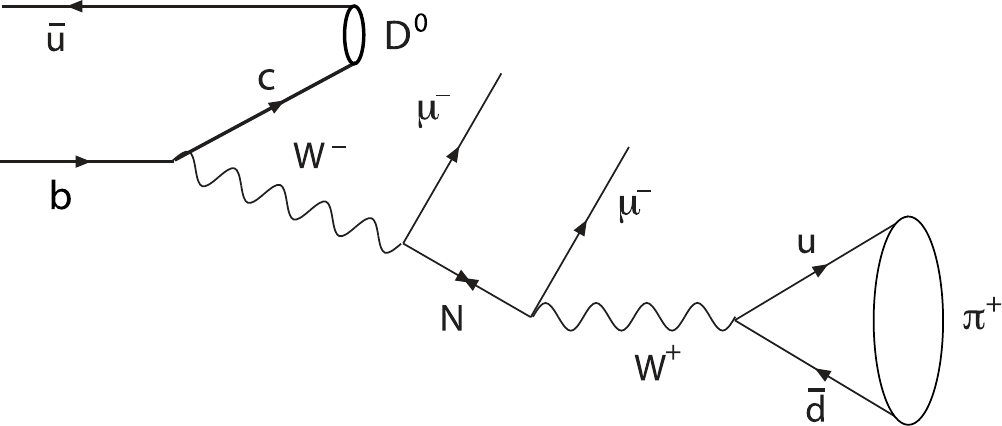}}
}
\caption{Lowest order Feyman diagram for the $\PB^{-} \to \Ppi^{+} \PD^{0} \Pmu^{-} \Pmu^{-}$ decay.}
\end{center}
\end{figure}
Figure 2 shows the four-body decay $\PB^{-} \to \Ppi^{+} \PD^{0} \Pmu^{-} \Pmu^{-}$, which was first analysed at the LHCb\cite{lhcb_majorana1,lhcb_majorana2}. In this case the accessible mass of Majorana neutrinos is smaller, between 260 MeV and 3300 GeV, but the rate is enhanced by $W$ coupling. \\

In particular, for LNV decays occuring via a fourth massive Majorana neutrino ${\nu_4}$, the observation of a LNV decay can provide not only information on the mass $m_{\nu_4}$ but also the  $W_{\nu4l}$ coupling strength $|V_{4l}|$. 
In absence of signal in all analysed LNV decays, limits on $|V_{4\mu}|$ coupling were set, which are presented in Figure 3.
\begin{figure}[h] 
\begin{center} 
\mbox{
    {\includegraphics[scale=0.4]{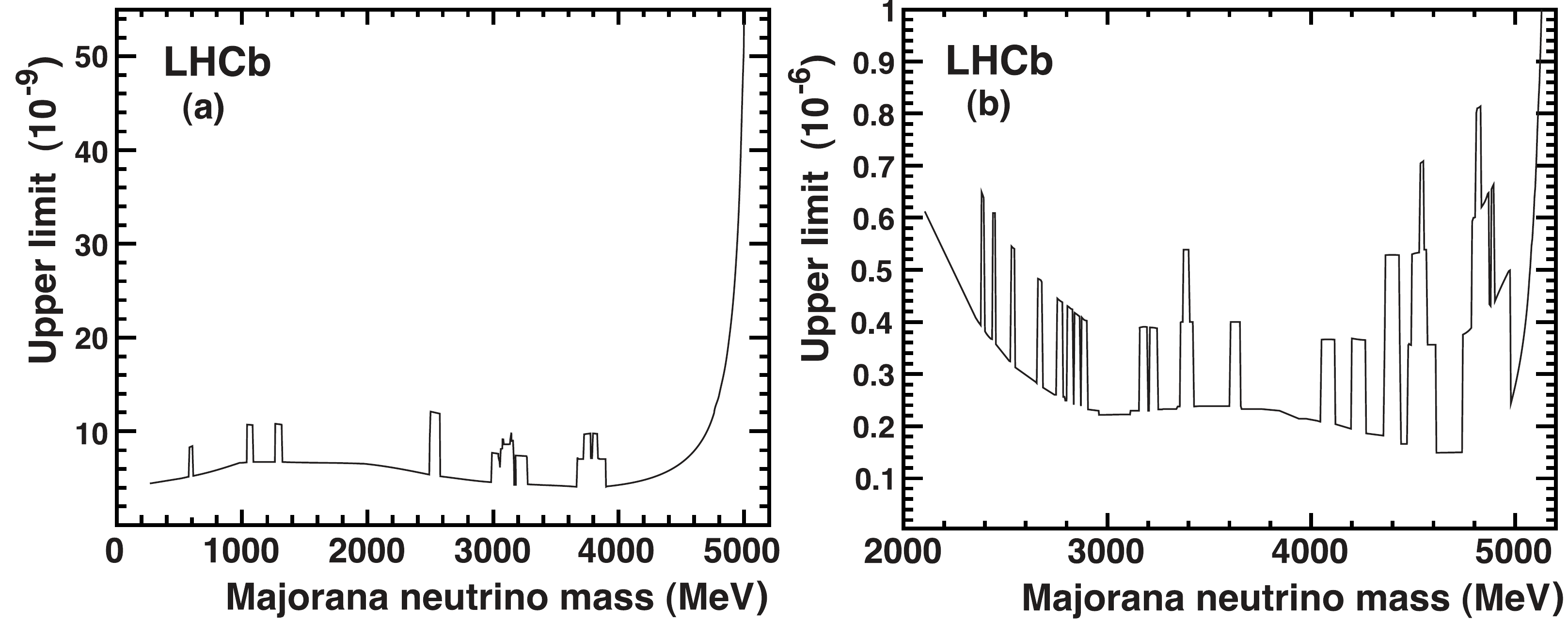}}

}
\caption{Exclusions on Majorana neutrino coupling as function of neutrino mass $m_{4}$, for (a) $\PB^- \to \Ppi^+ \mu^- \mu^-$ and (b) $\PB^- \to \PD^{+}_{s}  \mu^- \mu^-$}
\end{center}
\end{figure}
The current limits for various LNV decays from ${B}$ factories and LHCb are summarized in Table 1. 
\begin{table*}
\centering
\caption{Current limits on lepton number violating charm (a) and bottom (b) meson decays.}
\subfloat[][charm decays\label{tab:babar-charm}]{
\begin{tabular}{lclr}
channel & limit & & \\\hline
 $\mathcal{B}(\PDplus \to\Ppiminus \APelectron  \APelectron) $     &  $<1.9\times 10^{-6}$     &  @$90\,\%$ CL     &\cite{babar-charm} {\footnotesize{BaBar}} \\
 $\mathcal{B}(\PDplus \to\Ppiminus \APmuon      \APmuon)     $     &  $<2.0\times 10^{-6}$     &  @$90\,\%$ CL     &\cite{babar-charm} {\footnotesize{BaBar}} \\
 $\mathcal{B}(\PDplus \to\Ppiminus \APmuon      \APelectron) $     &  $<2.0\times 10^{-6}$     &  @$90\,\%$ CL     &\cite{babar-charm} {\footnotesize{BaBar}} \\
 $\mathcal{B}(\PDsplus\to\Ppiminus \APelectron  \APelectron) $     &  $<4.1\times 10^{-6}$     &  @$90\,\%$ CL     &\cite{babar-charm} {\footnotesize{BaBar}} \\
 $\mathcal{B}(\PDsplus\to\Ppiminus \APmuon      \APmuon)     $     &  $<14 \times 10^{-6}$     &  @$90\,\%$ CL     &\cite{babar-charm} {\footnotesize{BaBar}} \\
 $\mathcal{B}(\PDsplus\to\Ppiminus \APmuon      \APelectron) $     &  $<8.4\times 10^{-6}$     &  @$90\,\%$ CL     &\cite{babar-charm} {\footnotesize{BaBar}} \\
 $\mathcal{B}(\PDplus \to\PKminus  \APelectron  \APelectron) $     &  $<0.9\times 10^{-6}$     &  @$90\,\%$ CL     &\cite{babar-charm} {\footnotesize{BaBar}} \\
 $\mathcal{B}(\PDplus \to\PKminus  \APmuon      \APmuon)     $     &  $<10 \times 10^{-6}$     &  @$90\,\%$ CL     &\cite{babar-charm} {\footnotesize{BaBar}} \\
 $\mathcal{B}(\PDplus \to\PKminus  \APmuon      \APelectron) $     &  $<1.9\times 10^{-6}$     &  @$90\,\%$ CL     &\cite{babar-charm} {\footnotesize{BaBar}} \\
 $\mathcal{B}(\PDsplus\to\PKminus  \APelectron  \APelectron) $     &  $<5.2\times 10^{-6}$     &  @$90\,\%$ CL     &\cite{babar-charm} {\footnotesize{BaBar}} \\
 $\mathcal{B}(\PDsplus\to\PKminus  \APmuon      \APmuon)     $     &  $<13 \times 10^{-6}$     &  @$90\,\%$ CL     &\cite{babar-charm} {\footnotesize{BaBar}} \\
 $\mathcal{B}(\PDsplus\to\PKminus  \APmuon      \APelectron) $     &  $<6.1\times 10^{-6}$     &  @$90\,\%$ CL     &\cite{babar-charm} {\footnotesize{BaBar}} \\
 $\mathcal{B}(\PLambdac\to\APproton\APelectron  \APelectron) $     &  $<2.7\times 10^{-6}$     &  @$90\,\%$ CL     &\cite{babar-charm} {\footnotesize{BaBar}} \\
 $\mathcal{B}(\PLambdac\to\APproton\APmuon      \APmuon)     $     &  $<9.4\times 10^{-6}$     &  @$90\,\%$ CL     &\cite{babar-charm} {\footnotesize{BaBar}} \\
 $\mathcal{B}(\PLambdac\to\APproton\APmuon      \APelectron) $     &  $<16 \times 10^{-6}$     &  @$90\,\%$ CL     &\cite{babar-charm} {\footnotesize{BaBar}} \\
\end{tabular}
}
\hspace{.05\textwidth}
\subfloat[][bottom decays\label{tab:BLNV}]{
\begin{tabular}{lclrl}
channel & limit & & \\\hline
 $\mathcal{B}(\PBminus\to\Ppi^{+}\Pelectron\Pelectron) $    &  $<2.3\times 10^{-8}$    &  @$90\,\%$ CL      &\cite{babar-Blnv}  &{\footnotesize{BaBar}} \\
 $\mathcal{B}(\PBminus\to\PK^{+}\Pelectron\Pelectron)  $    &  $<3.0\times 10^{-8}$    &  @$90\,\%$ CL      &\cite{babar-Blnv}  &{\footnotesize{BaBar}} \\
 $\mathcal{B}(\PBminus\to\PK^{*+}\Pelectron\Pelectron) $    &  $<2.8\times 10^{-6}$    &  @$90\,\%$ CL    & \cite{cleo-Blnv}    &{\footnotesize{CLEO}}  \\
 $\mathcal{B}(\PBminus\to\Prho^{+}\Pelectron\Pelectron)$    &  $<2.6\times 10^{-6}$    &  @$90\,\%$ CL    & \cite{cleo-Blnv}    &{\footnotesize{CLEO}}  \\
 $\mathcal{B}(\PBminus\to\PD^{+}\Pelectron\Pelectron)  $    &  $<2.6\times 10^{-6}$    &  @$90\,\%$ CL      & \cite{belle-Blnv} &{\footnotesize{Belle}} \\
 $\mathcal{B}(\PBminus\to\PD^{+}\Pelectron\Pmuon)      $    &  $<1.8\times 10^{-6}$    &  @$90\,\%$ CL      & \cite{belle-Blnv} &{\footnotesize{Belle}} \\
 $\mathcal{B}(\PBminus\to\Ppi^{+}\Pmuon\Pmuon)         $    &  $<1.3\times 10^{-8}$    &  @$95\,\%$ CL    & \cite{lhcb-Blnv2}   &{\footnotesize{LHCb}}  \\
 $\mathcal{B}(\PBminus\to\PK^{+}\Pmuon\Pmuon)          $    &  $<5.4\times 10^{-7}$    &  @$95\,\%$ CL    &\cite{lhcb-Blnv}     &{\footnotesize{LHCb}}  \\
 $\mathcal{B}(\PBminus\to\PK^{*+}\Pmuon\Pmuon)         $    &  $<4.4\times 10^{-6}$    &  @$90\,\%$ CL    & \cite{cleo-Blnv}    &{\footnotesize{CLEO}}  \\
 $\mathcal{B}(\PBminus\to\Prho^{+}\Pmuon\Pmuon)        $    &  $<5.0\times 10^{-6}$    &  @$90\,\%$ CL    &\cite{cleo-Blnv}     &{\footnotesize{CLEO}}  \\
 $\mathcal{B}(\PBminus\to\PD^{+}\Pmuon\Pmuon)          $    &  $<6.9\times 10^{-7}$    &  @$95\,\%$ CL    & \cite{lhcb-Blnv2}   &{\footnotesize{LHCb}}  \\
 $\mathcal{B}(\PBminus\to\PD^{*+}\Pmuon\Pmuon)         $    &  $<2.4\times 10^{-6}$    &  @$95\,\%$ CL    &  \cite{lhcb-Blnv2}  &{\footnotesize{LHCb}}  \\
 $\mathcal{B}(\PBminus\to\PDs^{+}\Pmuon\Pmuon)         $    &  $<5.8\times 10^{-7}$    &  @$95\,\%$ CL    &  \cite{lhcb-Blnv2}  &{\footnotesize{LHCb}}  \\
 $\mathcal{B}(\PBminus\to\PDzero\Ppiplus\Pmuon\Pmuon)  $    &  $<1.5\times 10^{-6}$    &  @$95\,\%$ CL    &  \cite{lhcb-Blnv2}  &{\footnotesize{LHCb}} 
\end{tabular}
}
\end{table*}


\section{LFV, LNV and BNV in $\Ptau$ sector}
\subsection{LFV in $\Ptau$ sector}
Studies of LFV in $\Ptau$ decays has been performed extensively at the $B$ factories due to their high efficiency and clean environment. The most studied channels in the ${B}$ factories are $\Ptau \to 3 \Pmu$ and $\Ptau \to \Pmu \gamma$; their current experimental limits are $3.3 \times 10^{-8}$ from BaBar \cite{tau23mu_babar} and $2.1 \times 10^{-8}$ from Belle \cite{tau23mu_belle}, and $4.4 \times 10^{-8 }$ from BaBar \cite{tau2mugammababar}  and  $4.5 \times 10^{-8 }$ from Belle \cite{tau2mugammabelle}, respectively (all at $90 \%$ CL). \\
We focus on the channel $\Ptau^{-} \to \mu^{+} \mu^{-} \mu^{-} $. In the SM it has a branching ratio smaller than $10^{-54}$. Historically, this channel was studied in the ${B}$ factories, where $\Ptau$ are produced in pairs in a clean environment ($e^+ e^- \to \Ptau^+\Ptau^-)$. Using the thrust axis, efficient geometry tag of the other $\tau$ can be performed thereby reducing the combinatorial background and providing direct measurement of the number of $\tau$ produced. At the LHCb, the dominant mode for $\tau$ production $(78\%)$ is the leptonic decay $\PD^-_s \to \Ptau^- \bar{\nu}_\tau$ and $\Ptau$ tagging is not possible. This poses big experimental challenges to searches for decays like $\Ptau^- \to \Pmu^-\Pmu^-\Pmu^+$ at the LHCb. Nevertheless, the inclusive $\Ptau$ cross section of $79.5\pm 8.3 \enskip \mu$b is large at the LHCb compared to $0.919$ nb at the $B$ factories; this means two orders of magnitude more of $\tau$ leptons are produced in the LHCb in one nominal year than in the entire run of the $B$ factory experiments. In addition, final state muons have clean detector signatures thus studies of $\tau^- \rightarrow \mu^-\mu^-\mu^+$ and similar decays are totally viable at the LHCb.
\subsection{Search Strategy}
The search is performed by excluding a region of $\pm 30$MeV/c$^2$ around the $\tau$ mass until all the analysis choices are finalized (blind analysis). After passing the trigger, events are selected using loose cuts based on the kinematics of the reconstructed particles. Candidate events are then classified in a three-dimensional likelihood space: two multivariate classifiers $\mathcal{M}_\mathrm{3body}$ and $\mathcal{M}_\mathrm{PID}$, and the invariant mass of the $\tau$ candidate. $\mathcal{M}_\mathrm{3body}$ distinguishes displaced 3-body decays from N$(\geq3)$-body and separate combinations of tracks from different vertices using the kinematic and geometrical properties of the $\tau$ candidate. $\mathcal{M}_\mathrm{PID}$ quantifies the compatibility of each of the three decay products with the muon hypothesis using information from the RICH detector, calorimeters and muon chambers. Both classifiers are trained on signal and inclusive $b \bar{b}$ and $c \bar{c}$ background MC, and calibrated with the control channels $D^+_s \rightarrow \phi (\mu \mu) \pi^+ $ and $J/ \psi \rightarrow \mu \mu$ for $\mathcal{M}_\mathrm{3body}$ and $\mathcal{M}_\mathrm{PID}$ respectively. The space of each classifier is then binned; the number and boundaries of the bins are optimized using CLs\cite{cls} method. In both cases the optimal number of bins has been found to be 5. For the invariant mass classification, the signal mass window of $\pm 15$MeV/c$^2$ the expected $\tau$ mass is divided into 6 equally spaced bins and the signal shape is taken from the fit to $D^+_s \rightarrow \phi (\mu \mu) \pi^+ $. Both the central value of the mass window and the mass resolution are then corrected using the measured scaling and resolution at the LHCb.
\begin{figure}[h] 
\begin{center} 
\mbox{
	\subfloat[{Distribution for simulated background and the simulated signal as a function of the PID classifier.}]{\includegraphics[scale=0.33]{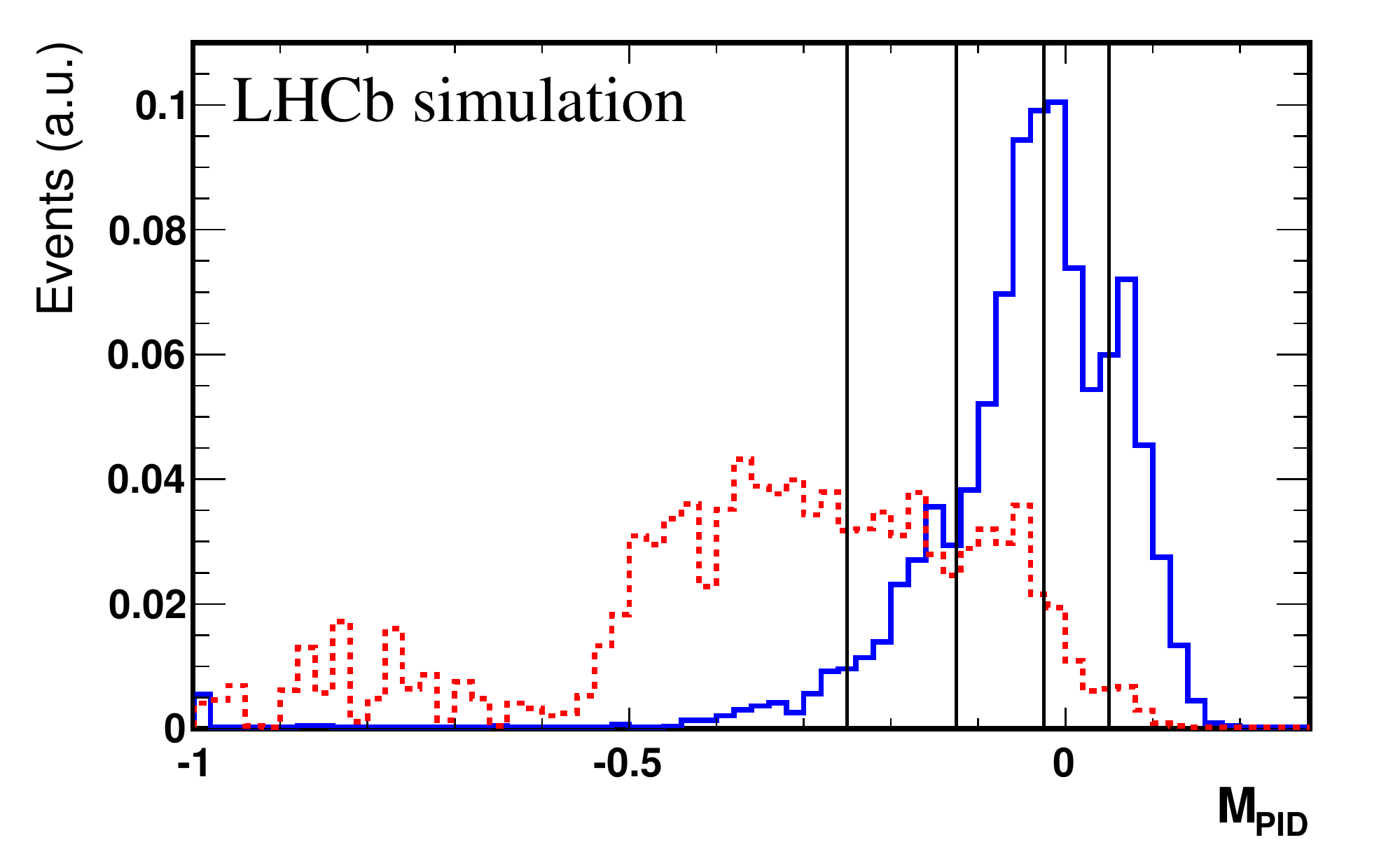}}
	\hspace{.05\textwidth}
    \subfloat[{Distribution for simulated background and the simulated signal as a function of the 3 body classifier.}]{\includegraphics[scale=0.33]{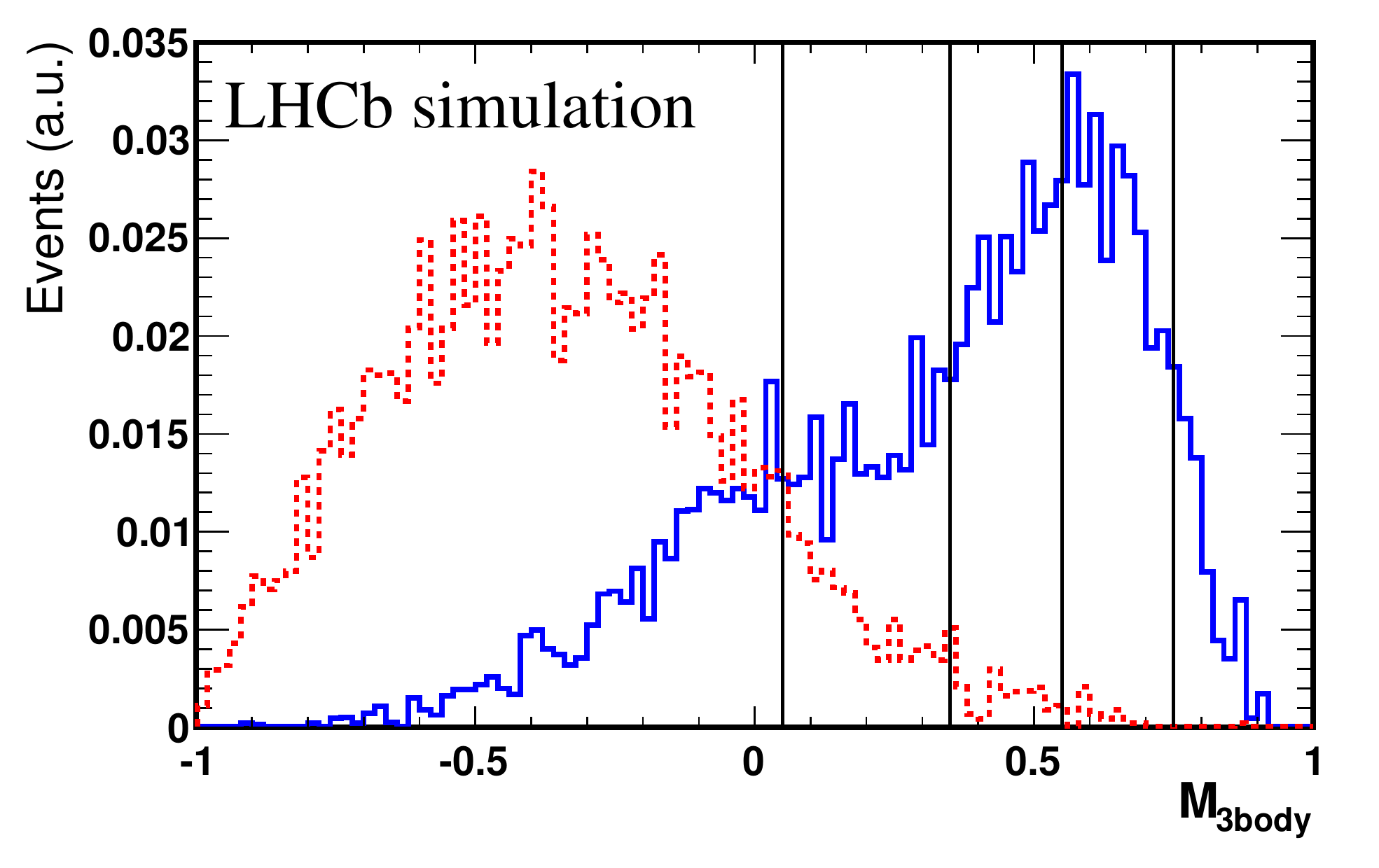}}
}
\caption{Distribution of signal events in the two multivariate likelihoods for signal (blue / solid) and background (red / dashed)}
\end{center}
\end{figure}
Figure 4 shows the distribution of the background and $\Ptau^- \to \Pmu^-\Pmu^-\Pmu^+$ signal MC candidates in each classifier, along with the binning. The vertical lines in Figure 4 are the bin boundaries. To calculate the branching ratio, the number of observed signal events is normalized to the number of events in the calibration channel $D^-_s \rightarrow \phi (\mu^+ \mu^-) \pi^- $
\begin{align*}\mathcal{B}(\Ptauon\to\Pmuon\Pmuon\APmuon)&=
\frac{\mathcal{B}(\PDsplus\to\Pphi(\APmuon\Pmuon)\Ppiplus)}{\mathcal{B}(\PDsplus\to\APtauon\Pnut)}\times f(\PDs)
\times\frac{\varepsilon_\text{norm}}{\varepsilon_\text{sig}}\frac{N_{sig}}{N_{Norm}}\enskip,\end{align*}
assuming negligible contribution from non-resonant events as suggested by data. $f(D_s)$ is the fraction of $\tau^-$ produced from ${D_s}$ decays; this factor is required since not all $\tau$ leptons are produced from $D_s^- \rightarrow \tau^- \bar{\nu}_{\tau}$; it is determined using the $b\bar{b}$ and $c \bar{c}$ cross sections and the inclusive $b \rightarrow \tau^-$ and $c \rightarrow \tau^-$. $\epsilon_{sig}$ and $\epsilon_{norm}$ denote the total efficiencies for the signal and normalization channels, which takes generation, selection and trigger efficiencies into account. 
\begin{figure}[h] 
\begin{center} 
\mbox{

	\subfloat[{Fit to observed events for $\Ptau^- \to \mu^- \mu^- \Pproton$ }]{\includegraphics[scale=0.25]{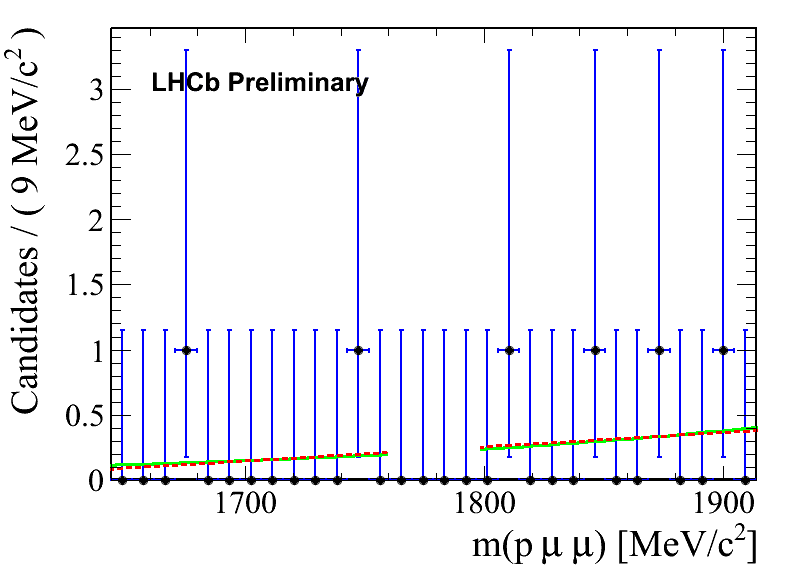}}
	\hspace{.05\textwidth}
    \subfloat[{Fit to observed events for $\Ptau^- \to \mu^+ \mu^- \APproton$ }]{\includegraphics[scale=0.25]{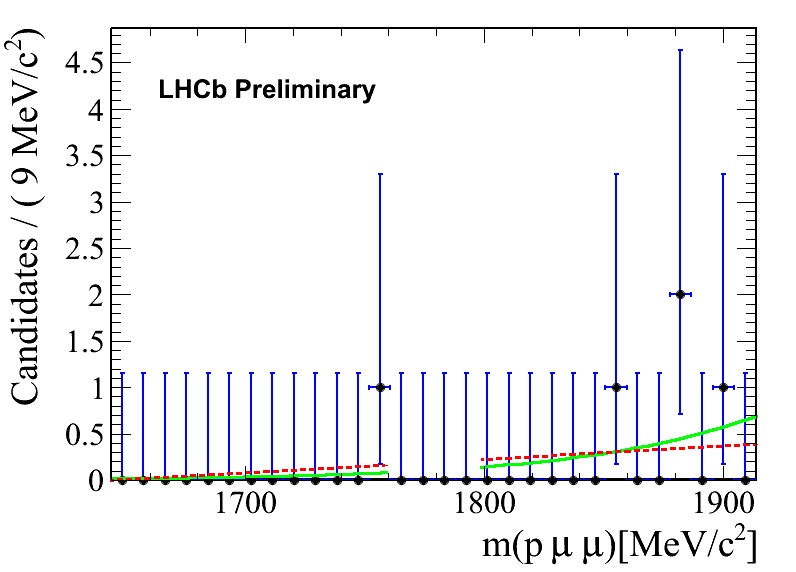}}
}
\mbox{

	\subfloat[{Fit to observed events for $\Ptau^- \to \mu^- \mu^- \mu^-$}]{\includegraphics[scale=0.3]{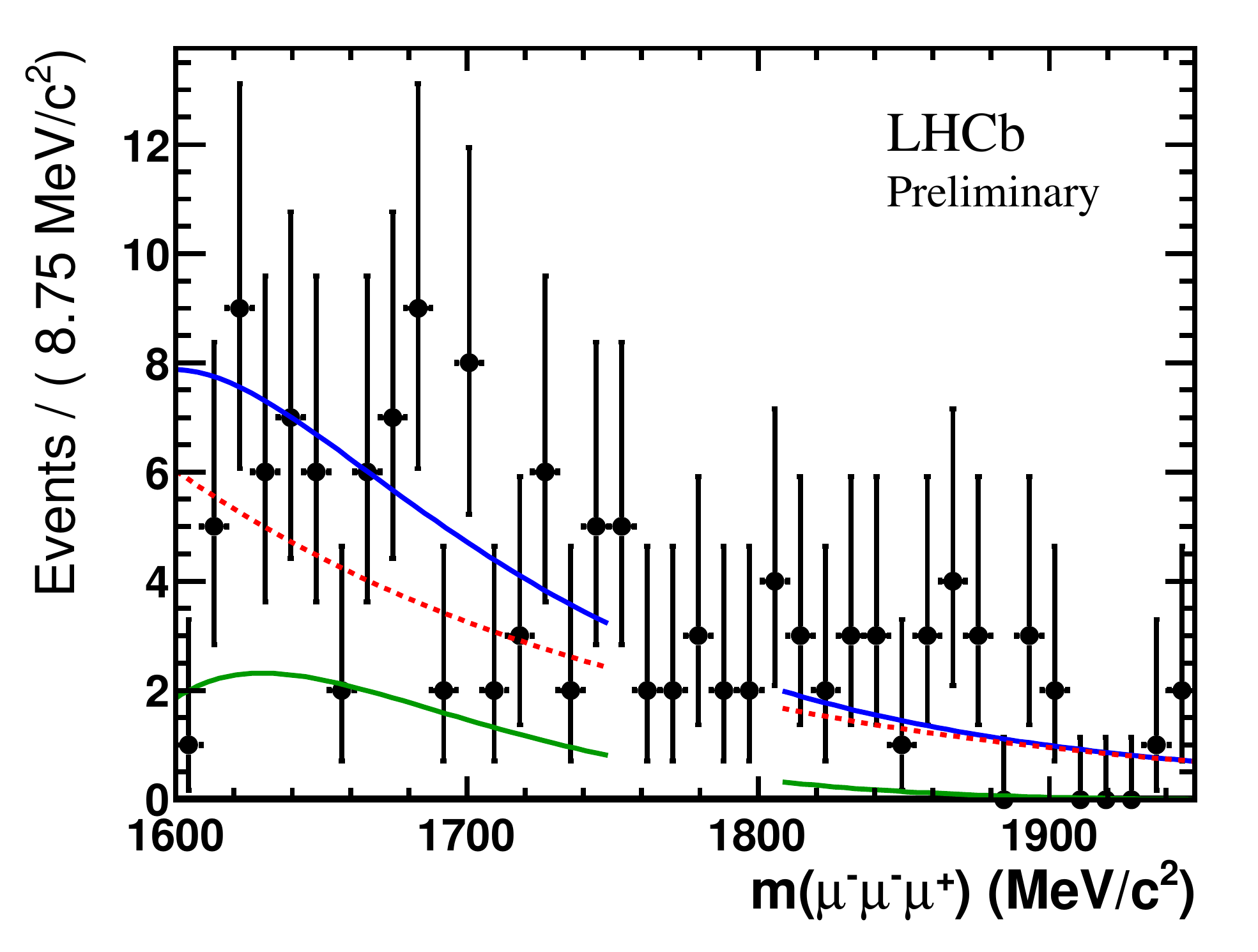}}
	
}
\caption{Fit to the events observed in highest likelihood bins for $\Ptau \to \mu \mu \mu$ and $\Ptau \to \mu \mu \Pproton$}

\end{center}
\end{figure}

\subsection{LNV and BNV in $\Ptau$ Sector}
We focus on the decays $\tau^- \rightarrow \bar{p} \mu^-\mu^+$ and $\tau^- \rightarrow {p} \mu^-\mu^-$ \cite{tau2pmumu_lhcb}. Both decays have $|B-L| =0$ which is predicted by many NP models. The analysis for these channels follow closely that of the $\Ptau^- \to \mu^- \mu^- \mu^-$ mode as described above. The main difference here is instead of PID BDT, hard PID cuts are applied on the muon and proton delta log likelihoods, which are optimized on signal MC and outer data side bands. Due to this hard cut, the normalization factor $ N_\mathrm{norm}$ is larger.
\section{Results}
The expected number of background events in the signal mass region is determined by interpolating from the invariant mass sidebands regions. In case of $\Ptau \to \mu \mu \mu$ the peaking SM background $\PD^+_s \to \eta(\mu\mu\gamma) \mu^+ \nu_\mu$  is taken into account by the fit, as this has been found to be the most relevant exclusive background. Combinatorial background is modelled in both cases with an exponential function. The  shapes of the invariant mass and the multivariate classifier response of the
 $\PD^+_s \to \eta(\mu\mu\gamma) \mu^{+} \nu_\mu$  decay are determined with a Monte Carlo sample corresponding to $5 fb^{-1}$ of data. For the $\Ptau \to \Pproton \mu \mu$ analysis, the expected number of background events is determined by interpolating from the mass sidebands also with an exponential and linear function. The difference is taken as systematic uncertainty. In the absence of signal an upper limit is calculated using CLs method. The results are listed in Table 2.

\begin{table}[tbh]
\centering
\caption{Limits on the branching fraction obtained by LHCb.}
\label{tab:tau23muresult}
\begin{tabular}{lclr}
Decay & Limit & CL \\\hline
$\Ptau^- \to \mu^- \mu^- \mu^-$             & $<6.3 \times 10^{-8}$ & @$90\,\%$ CL & \cite{tau23mu_lhcb} \\
$\Ptau^- \to \mu^+ \mu^- \APproton$                                     & $<3.4 \times 10^{-7}$ & @$90\,\%$ CL & \cite{tau2pmumu_lhcb} \\
$\Ptau^- \to \mu^- \mu^- \Pproton$                                    & $<4.4 \times 10^{-7}$ & @$90\,\%$ CL & \cite{tau2pmumu_lhcb} 

\end{tabular}
\end{table}

\section{Conclusions}

The LHCb performed its first measurements of LFV $\Ptau$ decays. The present limit for the $\Ptau^- \to \mu^- \mu^- \mu^-$ channel from the LHCb is still a factor 3 less restrictive than the ones set by the ${B}$ factories; however after the LHCb upgrade the expected limit with full data sample is foreseen to be around $8 \times 10^{-9}$. The LHCb also performed the first searches for the LNV and BNV channels: $\Ptau^- \to \mu^- \mu^- \Pproton$, $\Ptau^- \to \mu^+ \mu^- \APproton$. 

\section{Acknowledgements}
This work is supported by the Diamond Grant funded by the Polish Ministry of Science and the  computing grant funded by the PL-Grid Infrastructure.


\end{document}

%% file: econfmacros.tex
\def\beq{\begin{equation}}
\def\eeq#1{\label{#1}\end{equation}}
\def\eeqn{\end{equation}}

\def\beqa{\begin{eqnarray}}
\def\eeqa#1{\label{#1}\end{eqnarray}}
\def\eeqan{\end{eqnarray}}

\let\bar=\overbar

\def\Dslash{\not{\hbox{\kern-4pt $D$}}}
\def\dslash{\not{\hbox{\kern-2pt $\del$}}}

\def\msb{{\bar{\ssstyle M \kern -1pt S}}}

